\RequirePackage{fix-cm}
\documentclass[smallextended]{svjour3}       
\smartqed  
\usepackage{graphicx}
\usepackage{mathtools}
\usepackage[utf8]{inputenc}

\begin{document}

\title{Tangible reduction in learning sample complexity with large classical samples and small quantum system
}


\author{{Wooyeong~Song} \and {Marcin~Wie\'{s}niak} \and {Nana~Liu} \and {Marcin~Paw\l owski} \and Jinhyoung~Lee \and {Jaewan~Kim} \and {Jeongho~Bang}}


\institute{W.~ Song
\at Department of Physics, Hanyang University, Seoul 04763, Korea
\at Center for Quantum Information, Korea Institute of Science and Technology (KIST), Seoul, 02792, Korea
\and
M.~Wie\'{s}niak
\at Institute of Theoretical Physics and Astrophysics, Faculty of Mathematics, Physics and Informatics, University of Gda\'{n}sk, 80-308 Gda\'{n}sk, Poland
\at International Centre for Theory of Quantum Technologies, University of Gda\'{n}sk, 80-308 Gda\'{n}sk, Poland \and 
N.~Liu 
\at Institute of Natural Science, Shanghai Jiao Tong University, Shanghai 200240, China 
\at Ministry of Education, Key Laboratory in Scientific and Engineering Computing, Shanghai Jiao Tong University, Shanghai 200240, China
\at University of Michigan-Shanghai Jiao Tong University Joint Institute, Shanghai 200240, China 
\and 
M.~Paw\l{}owski 
\at International Centre for Theory of Quantum Technologies, University of Gda\'{n}sk, 80-308 Gda\'{n}sk, Poland 
\and
${}^\star$J.~Lee (\email{hyoung@hanyang.ac.kr})
\at Department of Physics, Hanyang University, Seoul 04763, Korea 
\and
${}^\star$J.~Kim (\email{jaewan@kias.re.kr}) 
\at School of Computational Sciences, Korea Institute for Advanced Study, Seoul 02455, Korea
\and
${}^\star$J.~Bang (\email{jbang@etri.re.kr}) \at Electronics and Telecommunications Research Institute, Daejeon 34129, Korea \and
${}^\star$Correspondence and requests for materials should be addressed to the last three authors.
}

\date{Received: date / Accepted: date}

\maketitle

\newcommand{\bra}[1]{\left<#1\right|}
\newcommand{\ket}[1]{\left|#1\right>}
\newcommand{\abs}[1]{\left|#1\right|}
\newcommand{\expt}[1]{\left<#1\right>}
\newcommand{\braket}[2]{\left<{#1}|{#2}\right>}
\newcommand{\ketbra}[2]{\left|{#1}\right>\left<{#2}\right|}
\newcommand{\commt}[2]{\left[{#1},{#2}\right]}
\newcommand{\tr}[1]{\mbox{Tr}{#1}}

\newcommand{\identity}{1\!\!1}

\newcommand{\add}[1]{\textcolor{blue}{#1}}

\begin{abstract}
Quantum computation requires large classical datasets to be embedded into quantum states in order to exploit quantum parallelism. However, this embedding requires considerable resources in general. It would therefore be desirable to avoid it, if possible, for noisy intermediate-scale quantum (NISQ) implementation. Accordingly, we consider a classical-quantum hybrid architecture, which allows large classical input data, with a relatively small-scale quantum system. This hybrid architecture is used to implement a sampling oracle. It is shown that in the presence of noise in the hybrid oracle, the effects of internal noise can cancel each other out and thereby improve the query success rate. It is also shown that such an immunity of the hybrid oracle to noise directly and tangibly reduces the sample complexity in the framework of computational learning theory. This NISQ-compatible learning advantage is attributed to the oracle's ability to handle large input features.


\end{abstract}

\section{Introduction}

Many celebrated quantum algorithms have shown promise for the quantum computational speedup~\cite{Shor99,Grover97,Harrow09}. However, apart from requiring considerable computational resources needed for the main calculation, many of them involve high costs for introducing ``big'' classical data into quantum states, for example, by accessing a useful quantum gadget, called quantum random-access memory (QRAM). An efficient QRAM algorithm, named bucket-brigade algorithm, has been developed~\cite{Giovannetti08-1,Giovannetti08-2}. Nevertheless, despite its promise for usefulness, it has some caveats related to physical resources and its ability to correct errors. This has been argued by S. Aaronson in Ref.~\cite{Aaronson15}. Thus, it is still unclear whether the bucket-brigade QRAM can be a prototype. This issue is well known in quantum computation and quantum machine learning (QML) (see Ref.~\cite{Arunachalam15} and Chap.~$5$ of Ref.~\cite{Ciliberto2018} for additional details on this issue). Therefore, researchers in the field of quantum computation and/or QML are unlikely to decisively state that {\em tangible} quantum (learning) speedups can be achieved with noisy intermediate-scale quantum (NISQ) technologies~\cite{Preskill2018,Arute2019}, when a large superposition of data is required. In this context, it is now highly desirable to show the possibility of achieving NISQ-technology-based speedups, namely, without using excessively large data superposition or, equivalently, without accessing the bucket-brigade QRAM. 

Toward this end, one promising approach is to consider a classical-quantum hybrid architecture and identify the optimal interplay between classical and quantum strategies. Such approaches have received increasing attention, an example being the consideration of variational formulations~\cite{Peruzzo14,Mcclean16,Khoshaman18,Zhu18,Havlivcek19}. Here, we devise an intriguing type of hybrid architecture, in which the input data remain classical but a small-scale quantum system is employed. Such an architecture differs from those of other classical-quantum hybridization, however it will render the task suitable for the NISQ implementation without requiring an excessively large superposition of the inputs, or equivalently, the bucket-brigade QRAM~\cite{Yoo14,Lee19,Bang19-1}. The motivation and background of this study are similar to those of the recent works in Refs.~\cite{Dunjko18,Harrow2020}.

We apply our hybridization to a classification task, a fundamental problem in computation and machine learning. For this, we employ a classical-quantum hybrid oracle designed on the basis of our main idea and assume that the oracle generates noisy samples due to errors resulting from the use of erroneous (internal) quantum gates. A model with such erroneous gates is often referred to as a noisy query model and is integrated in realistic models~\cite{Buhrman07,Cross15}. We demonstrate both analytically and numerically that our hybrid oracle can exhibit a high probability of success for queries. This advantage is attributed to the high capability of the oracle to explore a wider space of solutions, and it naturally leads to a quantum learning advantage, namely, a reduction in the sample complexity, in a computational learning framework~\cite{Valiant84,Langley95}. Our hybridization architecture is applicable in the relevant context of NISQ machine learning~\cite{Havlivcek19}.

\section{Classification with noisy samples}

Classification is a fundamental computational problem that is defined as follows~\cite{Ambainis04,Childs13,Arunachalam17}. Consider a Boolean function $h^\star$ that maps $\mathbf{x} = x_1 x_2 \ldots x_n$ ($x_j \in \{0,1\}$ $\forall j=1,\ldots,n$) to $h^\star(\mathbf{x}) \in \{0,1\}$~\footnote{Here, we consider a binary classification, i.e., mapping $\{0,1\}^n \to \{0,1\}$.}. Here, an algorithm (or a learner) can invoke an oracle to prepare a (finite) set of samples; given the inputs $\mathbf{x}$, the oracle returns the corresponding outputs $h^{\star}(\mathbf{x})$ and generates a set of samples $T=\{ (\mathbf{x}, h^\star(\mathbf{x})) \}$. The task is to identify a hypothesis $h$ close to $h^\star$ from among a set, say $H$, of candidates while minimizing the number of invocations of the oracle, or equivalently, the size of samples $\abs{T}$. The oracle can be implemented by using the general form of the Boolean function~\cite{Gupta06}
\begin{eqnarray}
h^\star(\mathbf{x}) = a_0 \oplus a_1 x_1 \oplus a_2 x_2 \oplus a_3 x_1 x_2 \oplus \cdots \oplus a_{2^n -1} x_1 x_2 \ldots x_n,
\label{eq:univ_c}
\end{eqnarray}
where $a_k \in \{0, 1\}$ ($k=0,1,\ldots,2^n-1$) are known as the Reed-Muller coefficients. In other words, given $\mathbf{x}=x_1 x_2\cdots x_n$, the oracle outputs, which could be either ideal or noisy. In this study, we are interested in noisy outputs, namely, the case where the oracle provides invalid outputs, resulting in a noisy sample set $\tilde{T}$. Note that the classification problem for noisy samples has widely been studied in computational learning theory since real-world data are not clean, which makes the computation much harder~\cite{Angluin88,Angluin94}. 

Here, we can introduce a corresponding error model of the generation of $\tilde{T}$. For this, each coefficient $a_k$ is changed such that $a_k \to a_k \oplus 1$ with a certain probability $\eta_k \le \frac{1}{2}$, and from the changed coefficients, noisy samples $(\mathbf{x}, h^{\star}(\mathbf{x}) \oplus e)$, where $e$ is to be $0$ (or $1$) with probability (say) $P(\mathbf{x})$ (or $1-P(\mathbf{x})$), can be obtained. This situation is often referred to as classification error in a noisy query model of computation~\cite{Bshouty98}. Such a model is useful since Eq.~(\ref{eq:univ_c}) is a general form and each $a_k$ can be realized by a Toffoli gate conditioned by $\mathbf{x}$ channels in a circuit~\cite{Toffoli80,Younes2004} (this is shown later). In this case, the oracle's invalid outputs are provided by flipping (i.e., $0 \rightleftharpoons 1$) the bit signal with probability $\eta_k$ before or after applying the $k$th Toffoli gate, namely, $a_k$. This error model can indicate a situation where the probability $P(\mathbf{x})$ of a successful query decreases as the problem size $n$ increases, and it is well suited for determining the effect of noise in complexity-theoretic studies~\cite{Yoo14}.

\section{Classical-quantum hybrid oracle}

\begin{figure}[t]
\centering
\includegraphics[width=0.60\textwidth]{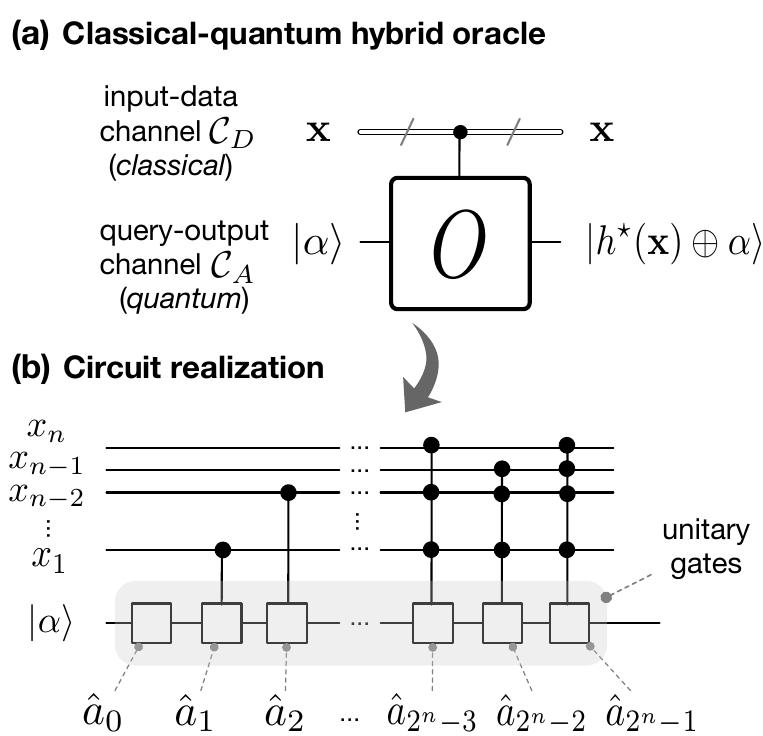}
\caption{\label{fig:oracle} (a) Schematic of our hybrid oracle. The oracle consists of two different input/output (I/O) channel types, one for input classical data $\mathbf{x} = x_1 x_2 \cdots x_n$ ($x_j \in \{0,1\}$ $\forall j=1,\ldots,n$), where $n$ can be very large, and the other for a single qubit, which produces the query-output states $\ket{h^\star(\mathbf{x})} \in \{0,1\}$. (b) Circuit realization of the oracle. This oracle applies $2^n$ unitary gates $\hat{a}_k \in \{ \hat{\sigma}_z, i\hat{\sigma}_y \}$ ($k=0,1,\ldots,2^n-1$) onto the qubit system, conditioned on the values of the classical bits $x_j$ in $\mathbf{x}$. In a fully classical case, these gates are either identity or logical-not. Here, we set $\alpha=0$ for convenience.}
\end{figure}

In classical studies, the oracle handles classical inputs and outputs. By contrast, (fully) quantum approaches begin by changing classical inputs into the corresponding quantum states, such as $\mathbf{x} \to \ket{\mathbf{x}}$, and the oracle performs the mapping 
\begin{eqnarray}
\sum_\mathbf{x}\ket{\mathbf{x},\alpha} \to \sum_\mathbf{x}\ket{\mathbf{x}, h^\star(\mathbf{x}) \oplus e},
\end{eqnarray}
in which the quantum parallelism resulting from the superposition of the samples is utilized~\footnote{Here we should clarify that such an oracle is employed for {\em data-sampling}, which it differs from those employed, for example, in the context of so-called amplitude amplification, where the oracle {\em marks} the relative phase on a single data state among superposed ones. In amplitude amplification studies, the primary objective is to reduce the number of {\em iterations} of the phase-marking oracle by using other incorporating modules~\cite{Van1998}.}. This oracle enables us to enjoy a quantum computational advantage in a specific situation~\cite{Arunachalam17}. However, such a process involves considerable cost and can even offset the advantages gained, as described at the beginning of this paper.

In line with our motivation and goal, we thus consider a classical-quantum hybrid oracle that allows classical inputs $\mathbf{x}$ and a small (a single qubit in our case) quantum system for the output (see Fig.~\ref{fig:oracle}(a)). More specifically, the oracle performs the mapping 
\begin{eqnarray}
\left( \mathbf{x}, \ket{\alpha} \right) \to \left( \mathbf{x} , \ket{\psi_\text{out}(\mathbf{x})} \right),
\end{eqnarray}
where $\ket{\alpha}$ is an arbitrary fiducial state and the output state $\ket{\psi_\text{out}(\mathbf{x})}$ has the form~\cite{Cross15,Bang19-1}
\begin{eqnarray}
\ket{\psi_\text{out}(\mathbf{x})} = \sqrt{P(\mathbf{x})} \ket{h^\star(\mathbf{x})} + \sqrt{Q(\mathbf{x})} \ket{h^\star(\mathbf{x}) \oplus 1},
\label{eq:query_output}
\end{eqnarray}
where $Q(\mathbf{x}) = 1-P(\mathbf{x})$.
Then, we can obtain a valid (or invalid) sample with probability $P(\mathbf{x})$ (or $Q(\mathbf{x})$), which is consistent with the error model described above. Note that $\mathbf{x}$ remains unaltered during and after the mapping. 

This hybrid oracle can be realized by using the circuit shown in Fig.~\ref{fig:oracle}(b). The circuit contains $2^n$ gates acting on the ancilla qubit, namely, the single-qubit gate $\hat{a}_0$ and $2^{n}-1$ of gates $\hat{a}_k$ ($k=1,2,\ldots,2^{n}-1$) conditioned on the classical bit values $x_1, x_2, \ldots, x_n$ in $\mathbf{x}$. The gates $\hat{a}_k$ are
\begin{eqnarray}
\hat{a}_k \in \left\{ \hat{\sigma}_z, i \hat{\sigma}_y \right\},~\text{for all}~k=0,1, \ldots, 2^n -1,
\label{eq:gates}
\end{eqnarray}
where $\hat{\sigma}_x$, $\hat{\sigma}_y$, and $\hat{\sigma}_z$ are the Pauli operators. Such a circuit realization is consistent with that described in the previous section and Eq.~(\ref{eq:univ_c}). Actually, each coefficient $a_k$ has a corresponding gate operation $\hat{a}_k$. More specifically, $a_k =0$ means that $\hat{a}_k$ leaves the bit signal unchanged (identity) and $a_k=1$ indicates that $\hat{a}_k$ flips the bit signal (logical-not). The oracle is thus characterized by a fixed set of $\hat{a}_k$'s. The gates $\hat{a}_k$ and their operation are not opened, and the task is to identify all of them. 

We can consider an error model consistently where the incorrect oracle answer $h^\star(\mathbf{x}) \oplus 1$ arises from the systematic errors in the gates $\hat{a}_k$. More specifically, the error can be described by $\ket{\psi_k(\mathbf{x})} \to \ket{\psi_k'(\mathbf{x})} = \hat{\epsilon}_k \ket{\psi_k(\mathbf{x})}$, where $\ket{\psi_k(\mathbf{x})}$ is the state passing through the gate $\hat{a}_k$. Here, $\hat{\epsilon}_k$ is a bit-flip operator defined as $\hat{\epsilon}_k = \sqrt{1-\eta_k}\hat{\identity} \pm i \sqrt{\eta_k}\hat{\sigma}_x$. In addition, we should consider another type of quantum error, the phase-flip error. This is a challenge for our hybrid oracle because in general, the phase-flip does not occur in classical gates. Thus, the prospect of quantum advantage (to be proved) can be claimed more confidently. The aforementioned error model is realistic, for example, for ion-trap or superconducting qubits, where the systematic errors are caused by imperfect control pulses~\cite{Debnath16}.

\section{Analysis}

We now analyze the query success probability $P_{C,Q}$ in Eq.~(\ref{eq:query_output}). Here, the subscripts $Q$ and $C$ refer to when the ancilla system in our hybrid oracle is quantum or classical, respectively. First, let us define a set $\Omega_\mathbf{x} = \{0, l_1, l_2, \ldots, l_{\kappa - 1} \}$ whose elements are the indices of the gates $\hat{a}_k$ that are ``activated'' (i.e., when the corresponding classical control bit $x_k=1$). The number of activated gates is given by $\kappa = 2^{\omega(\mathbf{x})}$, where $\omega(\mathbf{x})$ denotes the Hamming weight of $ \mathbf{x} = x_1 x_2 \cdots x_n$, namely, the number of $x_j$'s with a value of $1$ for $j \in [1,n]$. Then, $P_{C,Q}$ can be written in terms of $\omega$. In the classical case, $P_C(\omega)$ can be estimated as
\begin{eqnarray}
P_C(\omega) \simeq \sum_j^{\kappa/2} {{\kappa}\choose{2j}} \left(1-\overline{\eta}\right)^{2j} \overline{\eta}^{\kappa - 2j} \simeq \frac{1}{2}\left( 1 + e^{-\frac{2^\omega}{c}} \right),
\label{eq:Pc}
\end{eqnarray}
where $\overline{\eta}$ is the average error probability given by ${\overline{\eta} = \frac{1}{\abs{\Omega_\mathbf{x}}}\sum_{k \in \Omega_\mathbf{x}} \eta_k}$. The variance of the error probability, given by ${\Delta_\eta^2 = \frac{1}{\abs{\Omega_\mathbf{x}}}\sum_{k \in \Omega_\mathbf{x}} \eta_k^2 - \overline{\eta}^2}$, is assumed to be small. The factor $c$ is defined as 
\begin{eqnarray}
c = - \frac{1}{\ln{(1-2\overline{\eta})}} \simeq ({2\overline{\eta}})^{-1},
\end{eqnarray}
and is termed characteristic constant. Here, it is assumed that $O(\overline{\eta}^2) \to 0$. From Eq.~(\ref{eq:Pc}), $c$ can be interpreted as the characteristic number, say $\kappa$, of steps in the gate operations allowed before the oracle begins to give completely random outputs which cannot be used for learning.

In the case of our hybrid architecture, the success probability $P_Q(\omega)$ is expressed as
\begin{eqnarray}
P_Q(\omega) = \abs{\bra{h^\star(\mathbf{x})} \hat{\epsilon}_{l_{\kappa -1}}\hat{a}_{l_{\kappa -1}} \cdots \hat{\epsilon}_{l_1} \hat{a}_{l_1} \hat{\epsilon}_{0} \hat{a}_{0} \ket{\alpha}}^2.
\label{eq:Pq}
\end{eqnarray}
Here, using 
\begin{eqnarray}
\{ \hat{\sigma}_x, \hat{a}_k \}_{+} = \hat{\sigma}_x\hat{a}_k + \hat{a}_k\hat{\sigma}_x=0, 
\label{eq:prop_cal}
\end{eqnarray}
we can show that $P_Q(\omega)$ becomes unity in the limit $\Delta_\eta \to 0$. Thus, as long as the gate errors are regular, namely, $\eta_k = \overline{\eta}$ ($\forall k \in \Omega_\mathbf{x}$)~\cite{Cross15}, our hybrid oracle makes no mistakes. Evidently, our gates $\hat{a}_k$ in Eq.~(\ref{eq:gates}) satisfy the anticommutation relation in Eq.~(\ref{eq:prop_cal}), resulting in the amplitudes associated with gate errors {\em canceling out} through destructive interference.

However, it is impractical to realize such a perfect oracle that makes no mistakes, since in a realistic situation, it is difficult to meet the condition $\Delta_\eta = 0$. Furthermore, we should consider the phase-flip; this is crucial to generate a successful query output since the amplitudes changed by the gate errors would interfere in a disorderly manner because of the phase-flip. In fact, such a feature is often encountered in many physics models, for example, when dealing with localization problems~\cite{Eleuch17}. Consequently, $P_Q(\omega)$ has a form analogous to that of Eq.~(\ref{eq:Pc}). The characteristic constant $c$ is replaced with an ``effective'' one $c_\text{eff} \simeq (2\overline{\eta}_\text{eff})^{-1}$, where again, $O(\overline{\eta}^2) \to 0$. Here, $\overline{\eta}_\text{eff}$ is defined in terms of the effective average error of the $\hat{a}_k$'s. Notably, it is considerably smaller than $\overline{\eta}$, since $\overline{\eta}_\text{eff}$ originates from the errors remaining after destructive interference. Surprisingly, this feature, namely,
\begin{eqnarray} \label{eq:clarge}
\overline{\eta}_\text{eff} \le \overline{\eta}~\text{or equivalently}~c_\text{eff} \ge c,
\label{eq:eff_c}
\end{eqnarray}
does not depend on $\overline{\eta}$ but rather on $\Delta_\eta$. For a more detailed analysis, see Appendix~A.

On the basis of this feature, we show a quantum advantage. We begin with the average Hamming weight $\overline{\omega}=\frac{n}{2}$ for a given number $n$ of input bit strings. Then, on average, our hybrid oracle is useful up to the input bit-string length $n = 2\log_2{c_\text{eff}}$, while $n = 2\log_2{c}$ is the upper limit in the purely classical case. Hence, if $c_\text{eff} \ge c$, our hybrid oracle can be useful for larger bit-string inputs; this condition also implies the expansion of the space of legitimate samples that can be explored by the erroneous gates, which ranges from $O(e^{\left({2\overline{\eta}}\right)^{-2}\ln{2}})$ to $O(e^{\gamma^2\left({2\overline{\eta}}\right)^{-2}\ln{2}})$, where $\gamma = \frac{c_\text{eff}}{c}$ is larger than $1$. 

In addition to our theoretical analysis, we present numerical simulations in which $P_{C,Q}(\omega)$ are evaluated by counting a large number ($\simeq 10^5$) of queries for each given number of $\omega(\mathbf{x})$. Here, the identified data of $P_{C,Q}(\omega)$ are averaged over the trials ($\simeq 10^3$) again. To simulate a more realistic scenario, it is assumed that $\eta_k$'s are drawn from a normal distribution ${\cal N}(\overline{\eta}, \Delta_\eta)$. We also assume that the ancilla qubit undergoes the phase-flip errors on the gates $\hat{a}_k$ with probability $\chi_k \le \frac{1}{2}$, drawn from ${\cal N}(\overline{\chi}, \Delta_\chi)$. The simulation results confirm our theoretical analysis, allowing us to determine $c_\text{eff}$ and $\gamma$ for a given noise level. For example, when $\overline{\eta}=10^{-3}$ with $\Delta_\eta=5\%~\text{of}~\overline{\eta}$, our hybrid oracle would be applicable up to $n \simeq 27.23$ even in the presence of a phase-flip of $\overline{\chi}=10^{-2}$, whereas $n \simeq 17.93$ would be the limit in the purely classical case. Equivalently, the hybrid oracle can cover a space up to about $O(e^{1.09 \times 10^8})$ scale, which is considerably larger than the space, about $O(e^{1.73 \times 10^5})$ scale, allowed in the classical case. The determined $c_\text{eff}$ and $\gamma$ values are listed in Table~\ref{tab:eff_eta2}. For details on the methods and results of the numerical analysis, see Appendix~B.

\begin{table}[t]
\centering
\setlength{\tabcolsep}{0.25in}
\renewcommand{\arraystretch}{1.2}
\begin{tabular}{c  c  c}
\hline\hline
$\overline{\chi}$ & $c_\text{eff}$ (cf., $c=0.5 \times10^{3}$) & $\gamma = \frac{c_\text{eff}}{c}$ \\
\hline
no phase-flip &  $0.5 \times 10^{6.23}$ & $10^{3.23}$ \\ 
$10^{-4}$  &      $0.5 \times 10^{6.01}$ & $10^{3.01}$ \\
$10^{-3}$  &      $0.5 \times 10^{5.34}$ & $10^{2.34}$ \\ 
$10^{-2}$  &      $0.5 \times 10^{4.40}$ & $10^{1.40}$ \\ 
\hline\hline
\end{tabular}
\caption{Values of $c_\text{eff}$ and $\gamma$ are listed for several cases: $\overline{\chi}=0$, $10^{-4}$, $10^{-3}$, and $10^{-2}$. Here, $\overline{\eta}$ is assumed to be $10^{-3}$ and $\Delta_\eta = 0.05\overline{\eta}$.}
\label{tab:eff_eta2}
\end{table}

\section{Reduction of sample complexity}

The quantum advantage described above can be applied to a computational model of learning, the so-called PAC learning model. In this model, a finite set of $h^\star$, say $H^\star$, is said to be $(\epsilon, \delta)$-PAC learnable and we call the learner (or learning algorithm employed) $(\epsilon, \delta)$-PAC learner, if $P[E(h, h^\star) \le \epsilon] \ge 1-\delta$ is satisfied for $h^\star \in H^\star$. Here, $E(h, h^\star)$ denotes an arbitrary error function that indicates how $h$ and $h^\star$ are different. Borrowing terms from sampling theory in statistics, $\epsilon$ and $1-\delta$ are defined in terms of the accuracy and confidence, respectively. One beauty of PAC learning is that for any $h^\star \in H^\star$, if a learner is allowed to use a finite size, say $M_b(\epsilon, \delta)$, of samples, then the learner can be a $(\epsilon, \delta)$-PAC learner; in other words, we can know whether $H^\star$ is learnable in terms of the required samples, independent of the learning algorithm used. Usually, $M_b(\epsilon, \delta)$ is referred to as sample complexity. For a given set of {\em noisy} samples $\tilde{T}=\{ (\mathbf{x}, m) \}$ where $m$ is either $h^\star(\mathbf{x})$ or $h^\star(\mathbf{x}) \oplus 1$, it is known that $M_b(\epsilon, \delta)$ is at most~\footnote{The determination of its optimal (i.e., necessary and sufficient) condition is central and long-standing interest in computational learning theory, but this aspect is outside the scope of our study~\cite{Hanneke2016}.}
\begin{eqnarray}
\frac{2}{\epsilon^2 (1-2\xi)^2} \ln{\left(\frac{2\abs{H}}{\delta}\right)},
\end{eqnarray}
where $\xi$ is the imperfection of the oracle, that is, the probability $\xi$ that the oracle provides an incorrect output. For more details about PAC learning, see Refs.~\cite{Valiant84,Angluin94}.

Our previous analysis on the improvement of the query success rate can be extended to the reduction of sample complexity by letting $\xi = 1 - \overline{P}_Q(n)$, where $\overline{P}_Q(n)$ is the average query success probability given by $\overline{P}_Q(n) = \frac{1}{2^{n}}\sum_{\omega=0}^n {{n}\choose{\omega}} P_Q(\omega)$. Note that $\overline{P}_Q(n)$ would decrease as the number of gates (i.e., $2^n$) increases, indicating that the oracle's reliability rapidly worsens as the problem size $n$ increases. This is reasonable and shows that the error model is well suited for realistic situations. Here, we introduce the quantity 
\begin{eqnarray}
A_Q = \frac{1}{\left(2 \overline{P}_Q(n) - 1\right)^2} = \left[ \sum_{j=0}^\infty \frac{(-1)^j}{j !}\left(\frac{1}{\gamma c}\right)^j \left( \frac{2^j+1}{2} \right)^n \right]^{-2},
\label{eq:factor_Aq1}
\end{eqnarray}
which is consistently proportional to the sample complexity consistently but does not depend on $\epsilon$, $\delta$, and $\abs{H}$. Subsequently, for the sake of comparison, we also consider the case where a purely classical oracle is employed; in this case, the classical counterpart to $A_Q$ is $A_C$, which is defined using $\overline{P}_C(n)= \frac{1}{2^{n}}\sum_{\omega=0}^n {{n}\choose{\omega}} P_C(\omega)$ instead of $\overline{P}_Q(n)$. Then, we can show the reduction of the sample complexity from Eq.~(\ref{eq:eff_c}) and Eq.~(\ref{eq:factor_Aq1}) for $A_Q \le A_C$. 

This result tells us that for a large $n$, our hybrid oracle allows us to define a ($\epsilon$, $\delta$)-PAC learner, whereas a fully-classical oracle cannot. What is remarkable is that the sample complexity reduction is achieved with classical samples without the need to embed them into quantum states, unlike other classical-quantum hybrid schemes. Nevertheless, we note that if $n$ is too large (roughly, when $n \gg 2\log_2{\gamma c}$), it is also impractical to define a legitimate PAC learner, even with the quantum learning advantage gained.

\begin{figure}[t]
\centering
\includegraphics[width=0.45\textwidth]{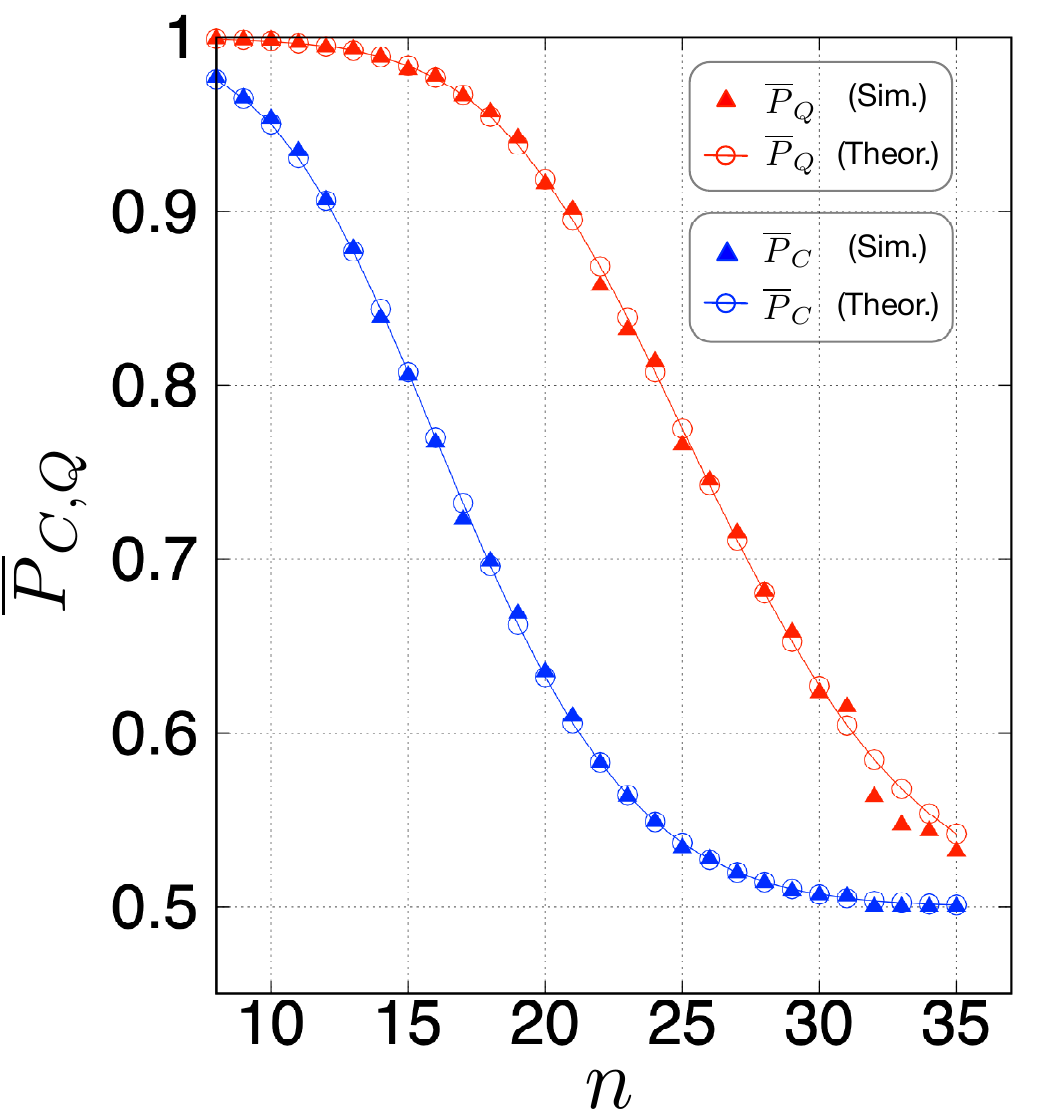}
\includegraphics[width=0.45\textwidth]{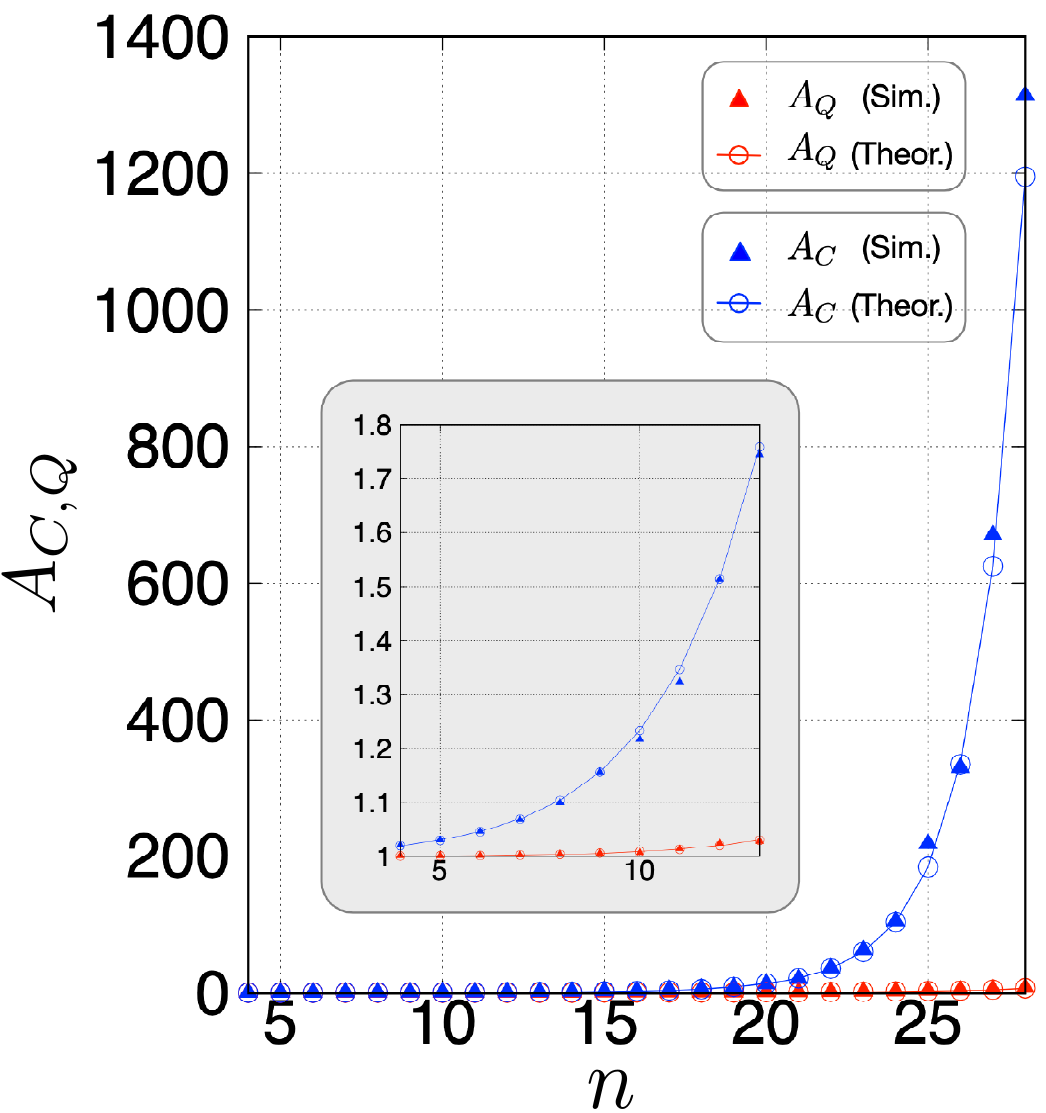}
\caption{\label{grp:results_ab} Numerical plots of $\bar{P}_{C,Q}$ (left) and $A_{C,Q}$ (right) with respect to $n$. Here, we have used $\overline{\eta}=10^{-3}$ with $\Delta_\eta = 0.05\overline{\eta}$ and $\overline{\chi}=10^{-2}$. Theoretical values are also presented for comparison. Refer to the main text for a detailed
description.}
\end{figure}

To confirm this feature, we perform numerical simulations. $\overline{P}_{C,Q}(n)$ are evaluated by repeating trials for randomly sampled inputs, with each bit $x_j$ ($j =1,2,\ldots,n$) in $\mathbf{x}$ being either $0$ or $1$ with probability $\frac{1}{2}$, and we examine the range $n=8$ to $n=35$. In these simulations, we assume $\overline{\eta}=10^{-3}$ with $\Delta_\eta = 0.05\overline{\eta}$ and $\overline{\chi}=10^{-2}$ with $\Delta_\chi = 0.1\overline{\chi}$. In Fig.~\ref{grp:results_ab}, we plot the dependence of $\overline{P}_{C,Q}$ and $A_{C,Q}$ on $n$. These results agree well with our theoretical predictions. Detailed descriptions of the method and analysis are provided in Appendix~C.

\section{Summary and remarks}
We have studied how a tangible quantum advantage can be achieved on the basis of a classical-quantum hybrid architecture distinct from other existing ones. Our approach is appealing, since the proposed hybridization architecture allows classical input data, without requiring (big) classical data to be embedded into a largely superposed quantum state, for example, by implementing QRAM. Instead, the quantum advantage is achieved with a small-sized quantum system (only a single-qubit is required in our case). We applied this approach to a binary classification problem and found that the oracle designed based on our main idea of classical-quantum hybridization, can improve the query success rate. This advantageous feature was attributed to the cancelation of the quantum amplitudes associated with the oracle's internal gate errors. Such (say) immunity could lead to a reduction of sample complexity in the PAC learning, facilitating the exploration of a large candidate space with noisy classical samples.

Of the two recent research directions in quantum machine learning---those seeking very strong complexity-theoretic evidence of quantum superiority to be realized in a full-scale quantum computer versus providing prospects for tangible quantum advantages which can be realized with NISQ technologies---our work pertains to the latter. We believe that the presented results are realizable and have potential to facilitate the development of an innovative classical-quantum hybrid technologies.

\begin{acknowledgements}
W.S., N.L., and J.B. are grateful to Gahyun Choi and Yonuk Chong for the valuable discussions on superconducting-qubit experiments. W.S., J.L., and J.B. acknowledge the financial support of the National Research Foundation of Korea (NRF) grants (No.~2019R1A2C2005504, No.~NRF-2019M3E4A1079666, and No.~2021M3E4A1038213), funded by the MSIP (Ministry of Science, ICT and Future Planning) of the Korea government. J.B. also acknowledge the support of the Institute of Information and Communications Technology Planning and Evaluation grant funded by the Korea government (Grant No.~2020-0-00890). W.S. and J.B. acknowledge the research project on developing quantum machine learning and quantum algorithm (No. 2018-104) by the ETRI affiliated research institute. W.S. acknowledge the KIST research program (2E31021). N. L. acknowledges funding from the Shanghai Pujiang Talent Grant (no. 20PJ1408400) and the NSFC International Young Scientists Project (no. 12050410230). N. L. is also supported by the Innovation Program of the Shanghai Municipal Education Commission (no. 2021-01-07-00-02-E00087), the Shanghai Municipal Science and Technology Major Project (2021SHZDZX0102) and the Natural Science Foundation of Shanghai grant 21ZR1431000. M.W. and M.P. acknowledge the ICTQT IRAP project of FNP (Contract No. 2018/MAB/5), financed by structural funds of EU. M.P. was supported under FNP grant First Team/2016-1/5. J.K. was supported in part by KIAS Advanced Research Program (No.~CG014604). J.B. was supported by a KIAS Individual Grant (No.~CG061003).
\end{acknowledgements}

\appendix{\bf Appendix A: Detailed calculations of $P_Q(\omega)$} \\

\begin{figure}[t]
\centering
\includegraphics[width=0.45\textwidth]{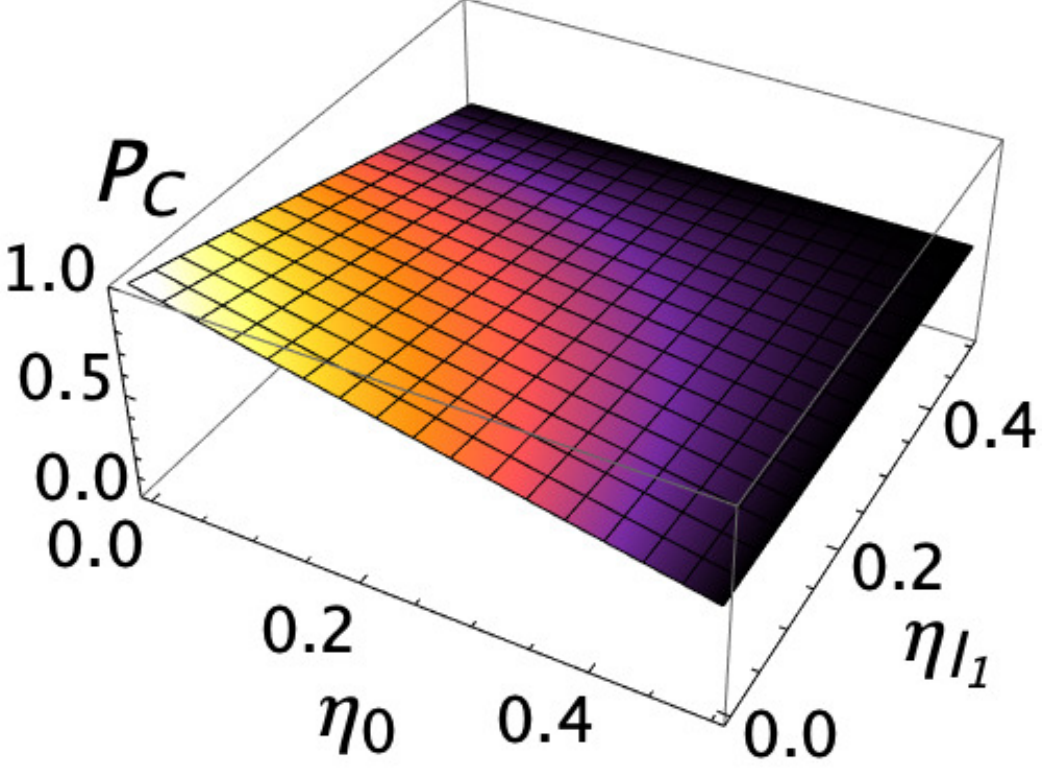}
\includegraphics[width=0.45\textwidth]{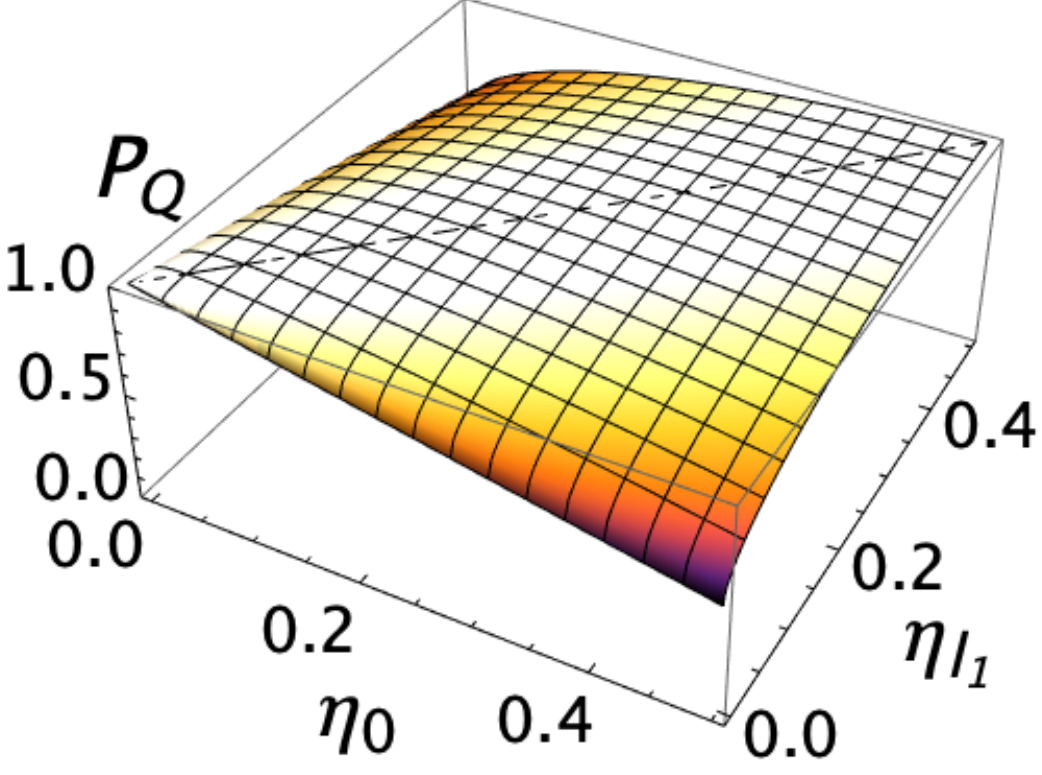}
\caption{\label{grp:Pcq_w1} 3D graphs of $P_{C}$ (left) and $P_{Q}$ (right) with respect to $\eta_0$ and $\eta_{l_1}$ for $\omega=1$. The advantage defined in Eq.~(\ref{eq:pq-pc}) is observed; $P_Q$ is always larger than $P_C$. Here, the most remarkable feature is that our hybrid oracle always yields correct results when $P_Q=1$ provided that $\eta_0 = \eta_{l_1}$.}
\end{figure}

Here, we present the procedure to calculate $P_Q(\omega)$ in Eq.~(6) of the main manuscript. We start by analyzing the simple case, i.e., of $\omega=1$. In particular, we consider an input $\mathbf{x} = x_1 x_2 \cdots x_n$ satisfying $x_{l_1} = 1$ for arbitrary $l_1 \in [1,n]$ and $x_{j}=0$ for all $j \neq l_1$. Subsequently, only two gates $\hat{a}_0$ and $\hat{a}_{l_1}$ are activated with $\Omega_\mathbf{x} = \{ 0, l_1 \}$. In a purely classical query, $P_C(\omega=1)$ is given as
\begin{eqnarray}
P_C(\omega=1)=\left(1- \eta_{l_1}\right)\left(1- \eta_0\right) + \eta_{l_1}\eta_{0},
\end{eqnarray}
where $\eta_k$ is the probability that a bit-flip error will occur at $\hat{a}_k$ ($k \in \{0, l_1\}$). Meanwhile, $P_Q(\omega=1)$ is calculated as below:
\begin{eqnarray}
P_Q(\omega=1) &=& \abs{\bra{h^\star(\mathbf{x})} {\hat{\epsilon}_{l_1} \hat{a}_{l_1} \hat{\epsilon}_{0} \hat{a}_{0}} \ket{\alpha}}^2, \nonumber \\
    &=& \abs{ \bra{h^\star(\mathbf{x})} \left( \sqrt{1-\eta_{l_1}}\hat{\identity} \pm i \sqrt{\eta_{l_1}}\hat{\sigma}_x \right) \hat{a}_{l_1} \left( \sqrt{1-\eta_0}\hat{\identity} \pm i \sqrt{\eta_0}\hat{\sigma}_x \right) \hat{a}_{0} \ket{\alpha} }^2, \nonumber \\
    &=& \left| \sqrt{1-\eta_{l_1}}\sqrt{1-\eta_0} \bra{h^\star(\mathbf{x})}\hat{a}_{l_1}\hat{a}_0\ket{\alpha} \pm i\sqrt{1-\eta_{l_1}}\sqrt{\eta_0} \bra{h^\star(\mathbf{x})}\hat{a}_{l_1}\hat{\sigma}_x\hat{a}_0\ket{\alpha} \right. \nonumber \\
    && \left. \pm i\sqrt{\eta_{l_1}}\sqrt{1-\eta_0} \bra{h^\star(\mathbf{x})}\hat{\sigma}_x\hat{a}_{l_1}\hat{a}_0\ket{\alpha} - \sqrt{\eta_{l_1}}\sqrt{\eta_0} \bra{h^\star(\mathbf{x})}\hat{\sigma}_x\hat{a}_{l_1}\hat{\sigma}_x\hat{a}_0\ket{\alpha} \right|^2, \nonumber \\
\label{eq:Pq_cal}
\end{eqnarray}
where $\hat{\epsilon}_k=\sqrt{1-\eta_k}\hat{\identity} \pm i \sqrt{\eta_k}\hat{\sigma}_x$ is the error operation, defined in the main manuscript. Using the properties in Eq.~(7) of the main manuscript, i.e., $\hat{\sigma}_x \hat{a}_k = -\hat{a}_k \hat{\sigma}_x$ and $\bra{h^\star(\mathbf{x})}\hat{a}_{l_1}\hat{a}_{0}\ket{\alpha}=1$, we can evaluate the following:
\begin{eqnarray}
\left\{
\begin{array}{l}
\bra{h^\star(\mathbf{x})}\hat{a}_{l_1}\hat{\sigma}_x\hat{a}_0\ket{\alpha} = -\bra{h^\star(\mathbf{x})}\hat{a}_{l_1}\hat{a}_0\hat{\sigma}_x\ket{\alpha}=0, \\ 
\bra{h^\star(\mathbf{x})}\hat{\sigma}_x\hat{a}_{l_1}\hat{a}_0\ket{\alpha} = \bra{h^\star(\mathbf{x})}\hat{a}_{l_1}\hat{a}_0\hat{\sigma}_x\ket{\alpha}=0, \\
\bra{h^\star(\mathbf{x})}\hat{\sigma}_x\hat{a}_{l_1}\hat{\sigma}_x\hat{a}_0\ket{\alpha} = -\bra{h^\star(\mathbf{x})}\hat{a}_{l_1}\hat{a}_0\ket{\alpha}=-1.
\end{array}
\right.
\label{eq:prop2}
\end{eqnarray}
Subsequently, using Eq.~(\ref{eq:prop2}), we can obtain
\begin{eqnarray}
P_Q(\omega=1) - P_C(\omega=1) = \Gamma_{0, l_1},
\label{eq:pq-pc}
\end{eqnarray}
where 
\begin{eqnarray}
\Gamma_{0, l_1}=2\sqrt{1-\eta_{l_1}}\sqrt{1-\eta_0}\sqrt{\eta_{l_1}}\sqrt{\eta_0} \ge 0.
\label{eq:Gamma}
\end{eqnarray}
This factor $\Gamma_{0, l_1}$ is from quantum superposition and clearly indicates the enhancement of the success probability with the condition $\Gamma_{0, l_1} \ge 0$. In Fig.~\ref{grp:Pcq_w1}, we depict the graphs of $P_{C,Q}$ with respect to $\eta_0$ and $\eta_{l_1}$. It is noteworthy that our hybrid oracle always yields correct results, i.e., $P_Q=1$, provided that $\eta_{l_1} = \eta_0$, even though $\eta_{l_1}$ and $\eta_{0}$ are large. This is the most remarkable feature in our classical--quantum hybrid query.

Subsequently, we consider the case of $\omega=2$, where a set of four gates, $\hat{a}_0$, $\hat{a}_{l_1}$, $\hat{a}_{l_2}$, and $\hat{a}_{l_3}$, are to be activated with $\Omega_\mathbf{x}=\{ 0, l_1, l_2, l_3 \}$. We subsequently calculate $P_Q(\omega=2)$ as follows:
\begin{eqnarray}
P_Q(\omega=2) = \abs{\bra{h^\star(\mathbf{x})} \hat{\epsilon}_{l_3} \hat{a}_{l_3} \hat{\epsilon}_{l_2} \hat{a}_{l_2} \hat{\epsilon}_{l_1} \hat{a}_{l_1} \hat{\epsilon}_{0} \hat{a}_{0} \ket{\alpha}}^2.
\label{eq:Pq(w=2)}
\end{eqnarray}
To proceed with the calculation, we introduce an identity $\hat{\identity}_{\beta,\beta^\perp}=\ket{\beta}\bra{\beta} + \ket{\beta^\perp}\bra{\beta^\perp}$, where the state $\ket{\beta}$ ($\beta \in \{0,1\}$) is defined with the following properties:
\begin{eqnarray}
\abs{\bra{h^\star(\mathbf{x})} \hat{a}_{l_3}\hat{a}_{l_2} \ket{\beta}}^2 = 1~\text{and}~\abs{\bra{\beta} \hat{a}_{l_1} \hat{a}_{0} \ket{\alpha}}^2 =1.
\end{eqnarray}
Using a mathematical method of substituting the identity $\hat{\identity}_{\beta,\beta^\perp}$ between $\hat{\epsilon}_{l_3} \hat{a}_{l_3} \hat{\epsilon}_{l_2} \hat{a}_{l_2}$ and $\hat{\epsilon}_{l_1} \hat{a}_{l_1} \hat{\epsilon}_{0} \hat{a}_{0}$ in Eq.~(\ref{eq:Pq(w=2)}), we can obtain
\begin{eqnarray}
P_Q(\omega=2) &=& \left| \left( \sqrt{1-\eta_{l_3}}\sqrt{1-\eta_{l_2}} + \sqrt{\eta_{l_3}}\sqrt{\eta_{l_2}} \right) \left( \sqrt{1-\eta_{l_1}}\sqrt{1-\eta_{0}} + \sqrt{\eta_{l_1}}\sqrt{\eta_{0}} \right) \right. \nonumber \\
 && - \left. \left( \sqrt{1-\eta_{l_3}}\sqrt{\eta_{l_2}} - \sqrt{\eta_{l_3}}\sqrt{1-\eta_{l_2}} \right) \left( \sqrt{1-\eta_{l_1}}\sqrt{\eta_{0}} - \sqrt{\eta_{l_1}}\sqrt{1-\eta_{0}} \right) \right|^2 \nonumber \\
\label{eq:Pq(w=2)-1}
\end{eqnarray}
Furthermore, after some algebraic simplifications, we can arrive at
\begin{eqnarray}
P_Q(\omega=2) - P_C(\omega=2) &=& \Gamma_{l_1, 0} \left( 1- 2\eta_{l_2}\right)\left( 1- 2\eta_{l_3}\right) + \Gamma_{l_2, l_1} \left( 1- 2\eta_{l_3}\right)\left( 1- 2\eta_{0}\right) \nonumber \\
    && + \Gamma_{l_3, l_2} \left( 1- 2\eta_{l_1}\right)\left( 1- 2\eta_{0}\right) + \Gamma_{0, l_3} \left( 1- 2\eta_{l_2}\right)\left( 1- 2\eta_{l_1}\right) \nonumber \\
    && - \Gamma_{l_2, 0} \left( 1- 2\eta_{l_3}\right)\left( 1- 2\eta_{l_1}\right) - \Gamma_{l_3, l_1} \left( 1- 2\eta_{l_2}\right)\left( 1- 2\eta_{0}\right) \nonumber \\
    && + 2 \Gamma_{l_3, l_2}\Gamma_{l_1, 0},
\label{eq:Pq(w=2)-2}
\end{eqnarray}
where 
\begin{eqnarray}
P_C(\omega = 2) &=& \left( 1 - \eta_{l_3} \right) \left( 1 - \eta_{l_2} \right) \left( 1 - \eta_{l_1} \right) \left( 1 - \eta_{0} \right) \nonumber \\
    && + \left( 1 - \eta_{l_3} \right) \left( 1 - \eta_{l_2} \right) \eta_{l_1} \eta_{0} + \left( 1 - \eta_{l_3} \right) \eta_{l_2} \left( 1 - \eta_{l_1} \right) \eta_{0} \nonumber \\
    && + \eta_{l_3} \left( 1 - \eta_{l_2} \right) \left( 1 - \eta_{l_1} \right) \eta_{0} + \left( 1 - \eta_{l_3} \right) \eta_{l_2} \eta_{l_1} \left( 1 - \eta_{0} \right) \nonumber \\
    && + \eta_{l_3} \left( 1 - \eta_{l_2} \right) \eta_{l_1} \left( 1 - \eta_{0} \right) + \eta_{l_3} \eta_{l_2} \left( 1 - \eta_{l_1} \right) \left( 1 - \eta_{0} \right) \nonumber \\
    && + \eta_{l_3} \eta_{l_2} \eta_{l_1} \eta_{0}.
\label{eq:Pc(w=2)}
\end{eqnarray}
Here, $\Gamma_{a, b}$ is defined as $\Gamma_{a,b}=2\sqrt{1-\eta_{a}}\sqrt{1-\eta_{b}}\sqrt{\eta_{a}}\sqrt{\eta_{b}}$ for $a \neq b \in \Omega_\mathbf{x} =\{ 0, l_1, l_2, l_3 \}$, similarly to Eq.~(\ref{eq:Gamma}). Subsequently, using Eq.~(\ref{eq:Pq(w=2)-1}) and Eq.~(\ref{eq:Pq(w=2)-2}), we demonstrate that the quantum advantage can be achieved with the positive factors $\Gamma_{a, b}$. Note that Eq.~(\ref{eq:Pq(w=2)-2}) could be negative thus exhibiting the disadvantage, e.g., when $\eta_{l_3}=\eta{l_1}=0$ or $\eta_{l_2}=\eta_{l_1}=0$ for all input $\mathbf{x}$. However, the aforementioned situation is not likely to occur in real physical systems. Consistent with the case of $\omega=1$, we observed that $P_Q(\omega=2)$ becomes unity when $\eta_{l_3}=\eta_{l_2}=\eta_{l_1}=\eta_{0}$.

By observing the two cases above, we can infer that the same method, i.e., of introducing the identities, can be used to calculate $P_Q(\omega)$ for arbitrary higher Hamming-weight inputs. The most remarkable construction, i.e., having unity query-success probability with equal error probabilities, can be generalized as well. Therefore, it can be sufficiently concluded that the enhancement in the query-success probability can be achieved for an arbitrary Hamming-weight in our hybrid query. \\

\appendix{\bf Appendix B: Numerical analyses with realistic conditions} \\

As mentioned in the main manuscript, in a more realistic situation, the amplitudes related to the errors are not completely canceled out owing to a nonzero $\Delta_\eta$, and $P_Q(\omega)$ exhibits an analogous form to $P_C(\omega)$ in Eq.~(6) of the main manuscript, with an ‘‘effective'' characteristic constant $c_\text{eff} \simeq (2\overline{\eta}_\text{eff})^{-1}$. Here, the effective average error $\overline{\eta}_\text{eff}$ is expected to be much smaller than $c$. This feature results in the quantum advantage that does not depend on the degree of $\overline{\eta}$ but only on $\Delta_\eta$, i.e., how ‘‘varying'' they are.

\begin{figure}[t]
\centering
\includegraphics[width=0.70\textwidth]{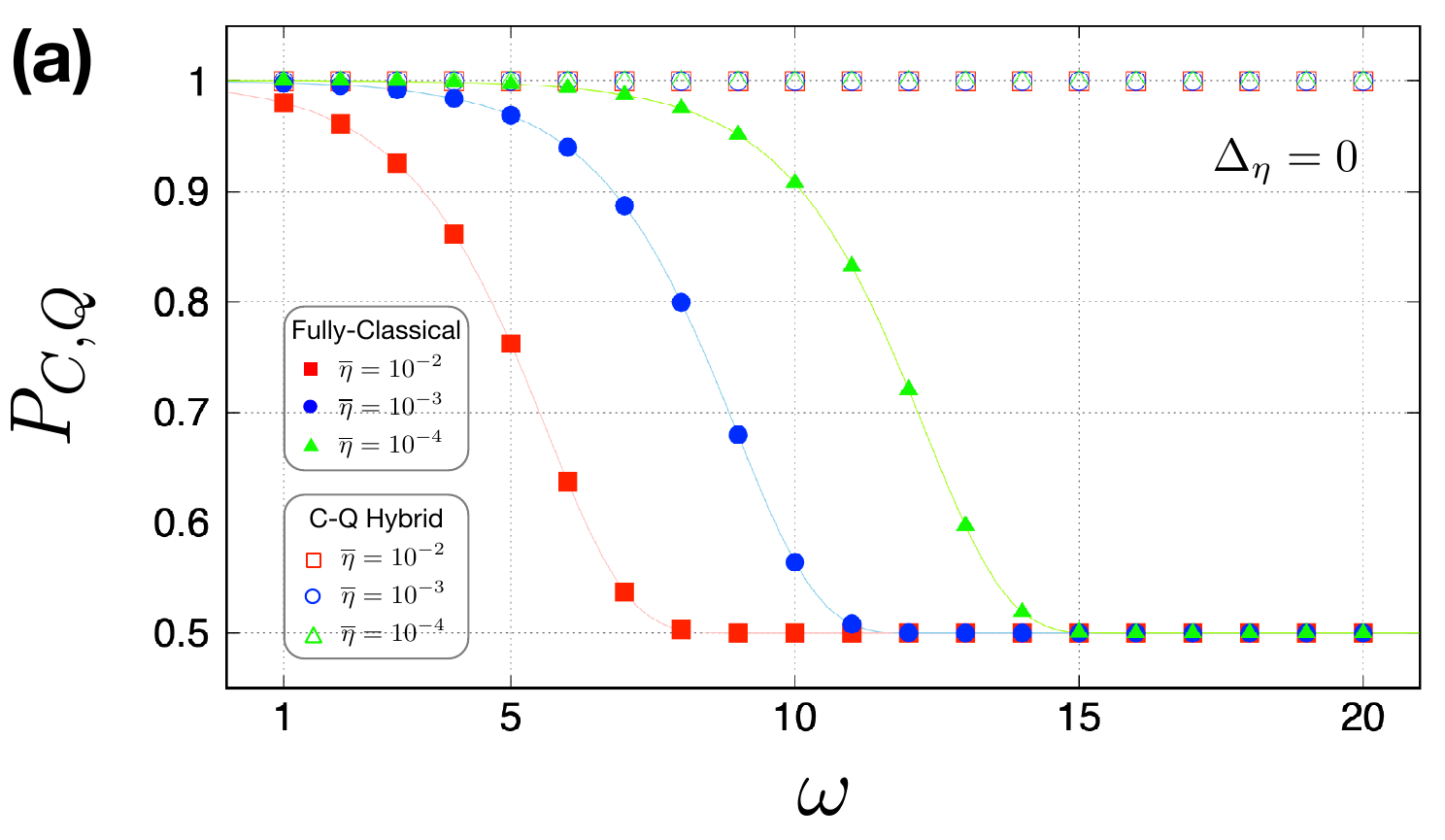} \\
\includegraphics[width=0.70\textwidth]{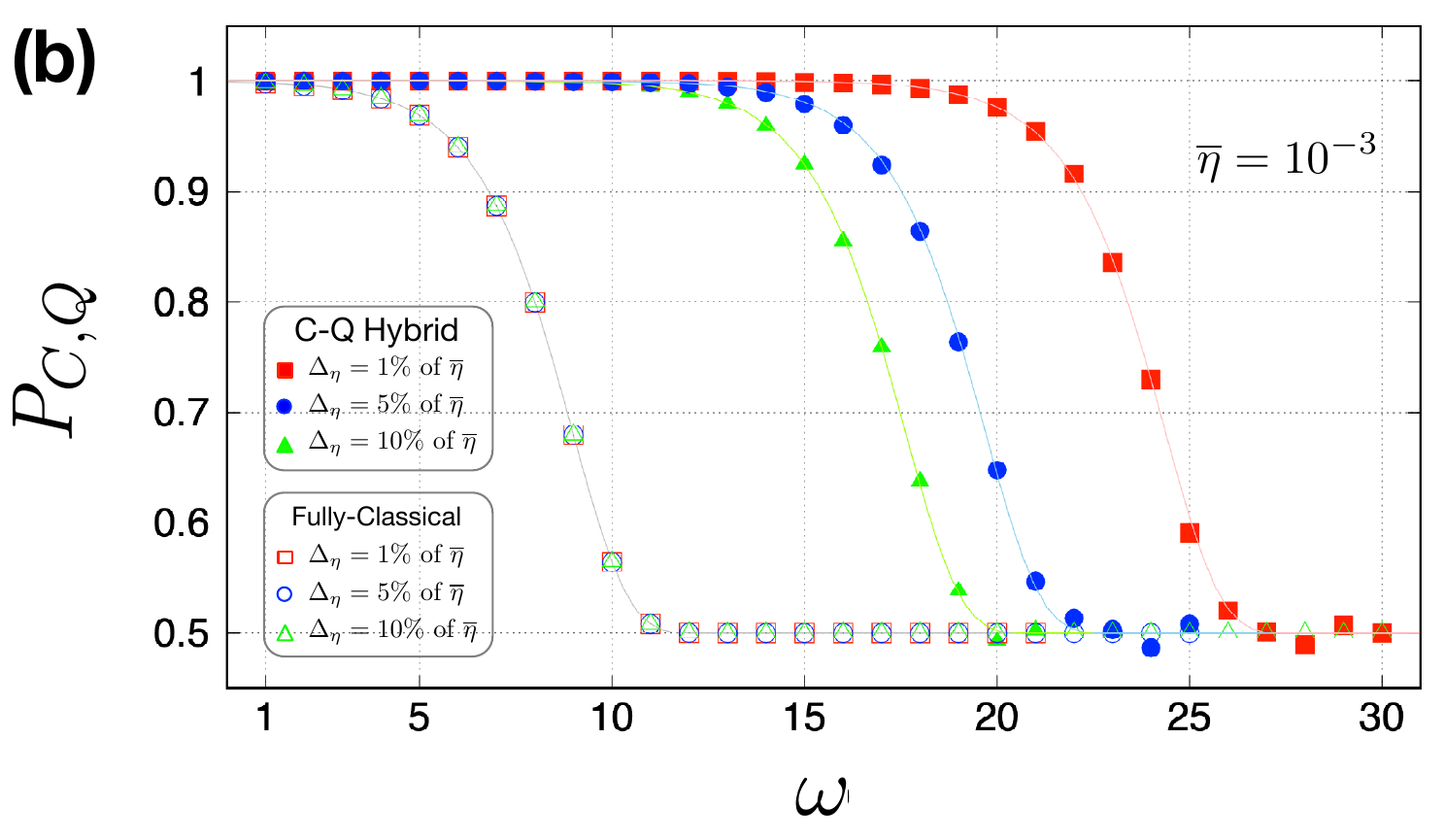}
\caption{\label{grp:bit-f} We plot the graphs of $P_{C,Q}$ versus $\omega$. The simulation is performed for randomly chosen inputs $\mathbf{x}$ and $h^\star$. In each simulation, $P_{C,Q}$ is evaluated by counting success and failure events over $10^5$ queries. One single data point of $P_{C,Q}$ is obtained by averaging $\simeq 10^3$ trials of the simulation. (a) First, we present the simulation data of $P_{C,Q}$ evaluated for $\overline{\eta} = 10^{-4}$, $10^{-3}$, and $10^{-2}$ with $\Delta_\eta=0$. The results show that $P_C$ rapidly approaches $\frac{1}{2}$ with increasing $\omega$, indicating good agreement with Eq.~(6) of our main manuscript (see red, blue, and green solid lines). Meanwhile, $P_Q$ remains unity for all the cases of $\overline{\eta}$, as predicted. (b) Next, we consider the realistic situation, assuming a normal distribution ${\cal N}(\overline{\eta}, \Delta_\eta)$. Here, we set $\overline{\eta}=10^{-3}$ with $\Delta_\eta = 1\%$, $5\%$, and $10\%~\text{of}~\overline{\eta}$. The data of $P_{C,Q}$ are shown to decay, but $P_Q$ is much slower. In such cases, the data $P_{Q}$ are well fitted by Eq.~(6) of the main manuscript with $c_\text{eff}$, indicating that the data agrees well with our theoretical predictions.}
\end{figure}

\begin{table}[t]
\centering
\setlength{\tabcolsep}{0.25in}
\renewcommand{\arraystretch}{1.2}
\begin{tabular}{c | c | c}
\hline\hline
$\Delta_\eta$ & $\overline{\eta}_\text{eff}$ (c.f., $\overline{\eta}=10^{-3}$) & $c_\text{eff}$ (c.f., $c=0.5 \times10^{3}$) \\
\hline
$1\%~\text{of}~\overline{\eta}$   & $\simeq 10^{-7.63}$ & $\simeq 0.5 \times 10^{7.63}$ \\ 
$5\%~\text{of}~\overline{\eta}$   & $\simeq 10^{-6.23}$ & $\simeq 0.5 \times 10^{6.23}$ \\
$10\%~\text{of}~\overline{\eta}$ & $\simeq 10^{-5.60}$ & $\simeq 0.5 \times 10^{5.60}$ \\
\hline\hline
\end{tabular}
\caption{Detailed values of $\overline{\eta}_\text{eff}$ and $c_\text{eff}$ for each $\Delta_\eta$.}
\label{tab:eff_eta1}
\end{table}

To corroborate and extend our theoretical predictions, we perform a numerical analysis. It starts with an input $\mathbf{x}$ of $\omega(\mathbf{x})$. We subsequently evaluate $P_{C,Q}(\mathbf{x})$ by counting the number of ``$h^\star(\mathbf{x})$'' (e.g.,, ``success'') and ``$h^\star(\mathbf{x}) \oplus 1$'' (e.g.,, ``failure''), such that $P_{C,Q}(\mathbf{x}) = {N_S}/\left({N_S + N_F}\right)$, where $N_S$ and $N_F$ denote the numbers of success and failure, respectively, and $N_S + N_F = 10^5$. Here, we use the Monte-Carlo approach to mimic quantum measurement statistics. This simulation is repeated for different values of $\eta_k$ (for $k \in \Omega_{\mathbf{x}}$) satisfying $c=(2\overline{\eta})^{-1}$. This condition enables us to analyze the data statistically (i.e., by averaging over the trials) without losing generality, even though in each simulation $\eta_k$ is changed with different $h^\star$. First, as an extreme but illustrative example, we consider the case of $\Delta_\eta=0$, i.e., by assuming $\eta_k = \overline{\eta}$ for all possible $k = 1, 2, \ldots, 2^n$. As results, we present the graphs of $P_{C,Q}$ versus $\omega$ as dots in Fig.~\ref{grp:bit-f}(a) for $\overline{\eta}=10^{-4}$, $10^{-3}$, and $10^{-2}$, where each data point of $P_{C,Q}$ is obtained by averaging over $\simeq 10^3$ trials. Here, it is observed that $P_C$ decays fast to $\frac{1}{2}$, indicating good agreement with Eq.~(6) of the main manuscript. The data of $P_Q$ are, meanwhile, shown to be unity without depending on the degree of $\overline{\eta}$, as predicted. Next, we consider a realistic situation, assuming that $\eta_k$ is drawn from a normal distribution ${\cal N}(\overline{\eta}, \Delta_\eta)$ for all $k = 1,2,\ldots,2^n$ (and hence for $k \in \Omega_{\mathbf{x}}$). Here, we set $\overline{\eta}=10^{-3}$ with $\Delta_\eta = 1\%$, $5\%$, and $10\%~\text{of}~\overline{\eta}$. The simulation results are shown in Fig.~\ref{grp:bit-f}(b). For all cases of $\Delta_\eta$, both $P_{C}$ and $P_{Q}$ decay to $\frac{1}{2}$; however, $P_Q$ is much slower. It is also observed that the data of $P_Q$ matched well with Eq.~(6) of the main manuscript, thus allowing us to identify the effective characteristic constant $c_\text{eff}$. The identified values of $c_\text{eff}$ and $\overline{\eta}_\text{eff}$ are listed in Tab.~\ref{tab:eff_eta1}; they manifest the predicted condition in Eq.~(8) of the main manuscript.

\begin{figure}[t]
\centering
\includegraphics[width=0.70\textwidth]{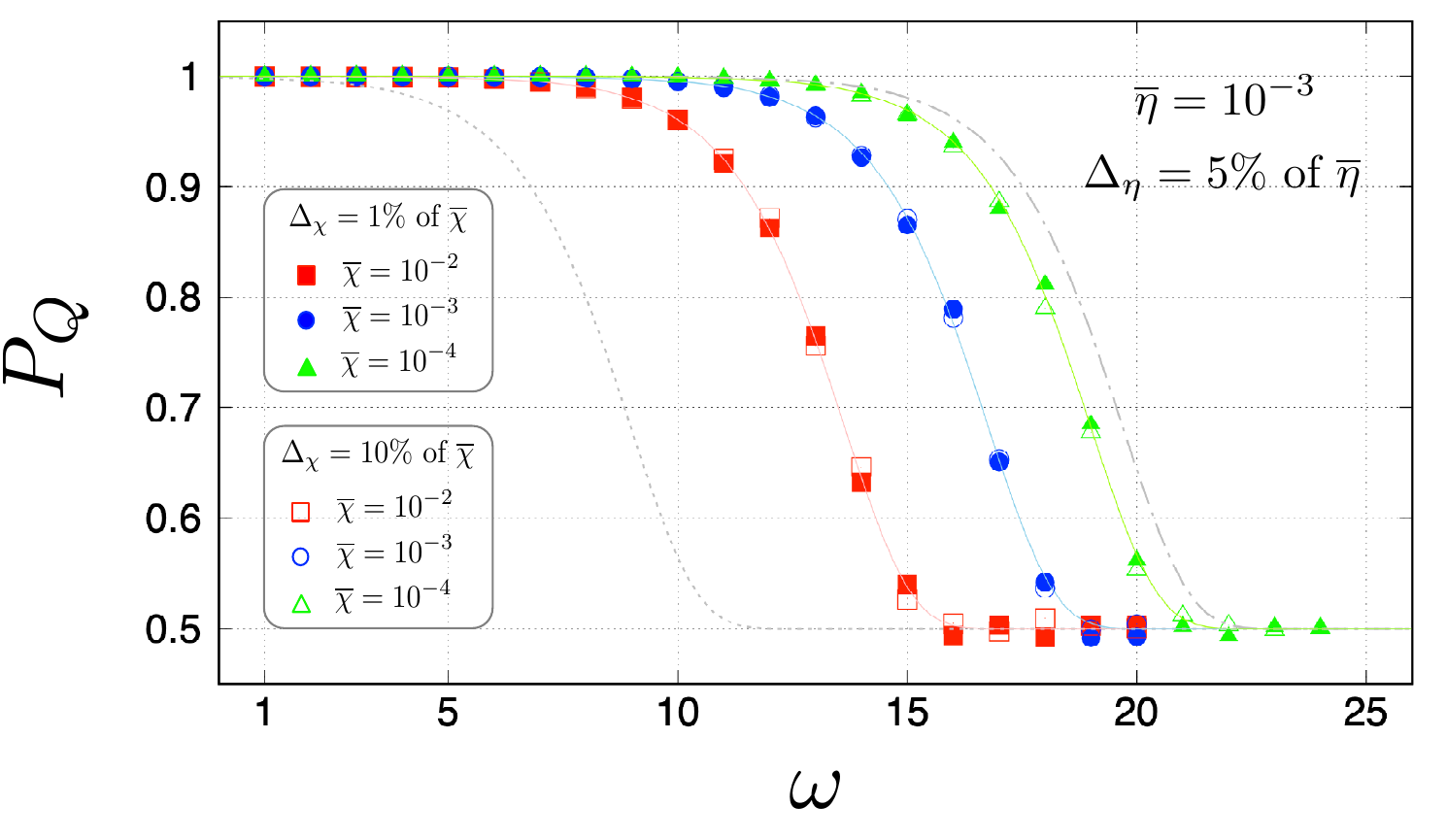}
\caption{\label{grp:phase-f} Graphs of $P_Q$ with respect to $\omega$ for $\overline{\chi}=10^{-4}$, $10^{-3}$, and $10^{-2}$. Each data point is obtained by averaging over $10^3$ simulations. The data fitted well to Eq.~(6) in the main manuscript, together with the parameter $c_\text{eff}$. The result shows that the quantum advantage becomes less pronounced as $\overline{\chi}$ is increased; however, it is still highly durable. It is noteworthy that the data obtained for $\Delta_\chi = 1\%~\text{of}~\overline{\chi}$ (filled square, circle, and triangle points) and $\Delta_\chi = 10\%~\text{of}~\overline{\chi}$ (empty square, circle, and triangle points) are almost identical (up to the order of $10^{-2}$); namely, $P_Q$ is not affected significantly by $\Delta_\chi$. The identified $\overline{\eta}_\text{eff}$ and $c_\text{eff}$ are listed in Tab.~\ref{tab:eff_eta1}.}
\end{figure}

\begin{table}[t]
\centering
\setlength{\tabcolsep}{0.25in}
\renewcommand{\arraystretch}{1.2}
\begin{tabular}{c | c | c}
\hline\hline
$\overline{\chi}$ & $\overline{\eta}_\text{eff}$ (c.f., $\overline{\eta}=10^{-3}$) & $c_\text{eff}$ (c.f., $c=0.5 \times10^{3}$) \\
\hline
no phase-flip &  $\simeq 10^{-6.23}$ & $\simeq 0.5 \times 10^{6.23}$ \\ 
$10^{-4}$  &      $\simeq 10^{-6.01}$ & $\simeq 0.5 \times 10^{6.01}$ \\
$10^{-3}$  &      $\simeq 10^{-5.34}$ & $\simeq 0.5 \times 10^{5.34}$ \\ 
$10^{-2}$  &      $\simeq 10^{-4.40}$ & $\simeq 0.5 \times 10^{4.40}$ \\ 
\hline\hline
\end{tabular}
\caption{Detailed values of $\overline{\eta}_\text{eff}$ and $c_\text{eff}$ for each $\overline{\chi}$.}
\label{tab:eff_eta2_appendix}
\end{table}

For a more realistic condition, we consider another type of error, i.e., phase-flip in the assistant qubit that would be crucial for maintaining a higher success rate of the query. In particular, we assume that the phase-flip errors primarily occur when the qubit travels between $\hat{a}_k$ and $\hat{a}_{k+1}$ with a certain probability $\chi_k \le \frac{1}{2}$. First, when $\Delta_\eta=0$ (or equivalently, $\eta_k = \overline{\eta}$) for all $k$, the phase-flip errors do not affect the query process and $P_{Q}$ becomes unity. In the realistic case, namely of $\Delta_\eta \neq 0$, however, it is predicted that the amplitudes of the bit-flip errors would interfere {\em disorderly} owing to the phase-flip, and eventually the quantum advantage becomes smaller, as described in our main manuscript. Thus, we perform the simulations and present the data of $P_Q$ in Fig.~\ref{grp:phase-f}. Here, $\chi_k$ is assumed to be drawn from ${\cal N}(\overline{\chi}, \Delta_\chi)$ for all $k=1,2,\ldots,2^n$. The simulation data are generated for $\overline{\chi}=10^{-4}$, $10^{-3}$ and $10^{-2}$. Here, we set $\overline{\eta}=10^{-3}$ with $\Delta_\eta = 5\%~\text{of}~\overline{\eta}$. The data are well fitted by Eq.~(6) of our main manuscript, and $c_\text{eff}$ are well estimated from the data (see Tab.~\ref{tab:eff_eta2_appendix}). As expected, the quantum advantage becomes less pronounced as $\overline{\chi}$ is increased; however, it can still exhibit a higher success rate of the query. It is noteworthy that the data obtained for both $\Delta_\chi = 1\%~\text{and}~10\%~\text{of}~\overline{\chi}$ are almost identical (up to the second digit of a decimal). \\


\appendix{\bf Appendix C: Reduction in learning sample complexity in the framework of probably-approximately-correct (PAC) learning} \\

In a probably-approximately-correct (PAC) learning model, a learner (or equivalently, a learning algorithm) samples a finite set of training data $\{ (\mathbf{x}_i, h^\star(\mathbf{x}_i)) \}$ ($i = 1,2,\ldots,M$) by accessing an oracle, aiming at obtaining the best hypothesis $h$ close to $h^\star$ for a given set, e.g., ${  H}$, of the hypothesis $h$. Here, $\mathbf{x}_i$ is typically assumed to be drawn uniformly. Subsequently, a learning algorithm is a ($\epsilon$, $\delta$)-PAC learner (under uniform distribution), if the algorithm obtains an $\epsilon$-approximated correct $h$ with probability $1-\delta$; more specifically, satisfying 
\begin{eqnarray}
P[E(h, h^\star) \le \epsilon] \ge 1-\delta,
\label{eq:pac_e}
\end{eqnarray}
where $E(h, h^\star)$ denotes the error. Here, if $h$ identified by the algorithm agrees with
\begin{eqnarray}
M \ge \frac{1}{\epsilon}\ln{\frac{{\abs{  H}}}{\delta}}
\label{eq:sample_C}
\end{eqnarray}
of samples constructed from the oracle, then Eq.~(\ref{eq:pac_e}) holds. Here, $\abs{H} \le 2^{2^n}$ denotes the cardinality of ${  H}$. Eq.~(\ref{eq:sample_C}) is known as the bound of the sample complexity~\cite{Valiant84,Langley95}, i.e., it yields the minimum number of training samples to successfully learn $h \in {  H}$ satisfying Eq.~(\ref{eq:pac_e}). Such a sample complexity bound derived from the previous studies can directly be carried over to our scenario; in our classical--quantum hybrid query scheme, the same sample complexity bound exists, because $\mathbf{x}_i$ and $h^\star(\mathbf{x}_i)$ identified by the measurement on $\ket{\psi_\text{out}(\mathbf{x}_i)}$ are classical.

However, in the case where the oracle is not perfect, the bound of sample complexity in Eq.~(\ref{eq:sample_C}) is modified as follows: First, we draw a sequence of the training data $\{ (\mathbf{x}_1, m_1), (\mathbf{x}_2, m_2), \ldots, (\mathbf{x}_M, m_M) \}$ sampled from our classical--quantum hybrid oracle, where $m_i \in \{ h^\star(\mathbf{x}_i), h^\star(\mathbf{x}_i) \oplus 1 \}$ denotes the outcome of the measurement performed on $\ket{\psi_\text{out}(\mathbf{x}_i)}$. Subsequently, if the sampling is performed with 
\begin{eqnarray}
M \ge 2 A_Q\ln{\left(\frac{2\abs{  H}}{\delta}\right)^{\frac{1}{\epsilon^2}}},
\label{eq:sc_q}
\end{eqnarray}
we can verify that Eq.~(\ref{eq:pac_e}) holds for the algorithm that obtains $h$ maximizing $\overline{P}_Q$. In fact, it has been proven that the additional factor $A_Q$ is given as~\cite{Angluin94}
\begin{eqnarray}
A_Q = \frac{1}{\left(2 \overline{P}_Q(n) - 1\right)^{2}}.
\label{eq:factor_Aq1}
\end{eqnarray}
It is noteworthy that in the purely classical case, the corresponding factor, e.g., $A_C$, is given with $\overline{P}_C$ instead of $\overline{P}_Q$. Thus, we can derive the reduction in the sample complexity with the condition $A_Q \le A_C$ from $\overline{P}_Q \ge \overline{P}_C$. To view this explicitly, we rewrite $A_Q$ in Eq.~(\ref{eq:factor_Aq1}) to a more useful form:
\begin{eqnarray}
A_Q = \left[ \frac{1}{2^n}\sum_{\omega=0}^n {{n}\choose{\omega}} e^{-\frac{2^\omega}{\gamma c}} \right]^{-2} = \left[ \sum_{j=0}^\infty \frac{(-1)^j}{j !}\left(\frac{1}{\gamma c}\right)^j \left( \frac{2^j+1}{2} \right)^n \right]^{-2}
\label{eq:factor_Aq2}
\end{eqnarray}
This implies that a small increment in the sample complexity bound when $n$ is small increases {\em abruptly} from near $n \simeq 2\log_2{\gamma c}$. As $A_C$ is characterized by $c$ without $\gamma$, we can interpret $\gamma$ as a quantum learning advantage in the PAC learning framework; i.e., for any large $n$, we can define a ($\epsilon$, $\delta$)-PAC learner with our hybrid oracle, unlike with a fully classical one. It is noteworthy that if $n$ is excessively large, i.e., when $n \gg 2\log_2{\gamma c}$, it is impractical to define a legitimate PAC learner even with our hybrid oracle. This result is consistent with the recent theoretical study in Ref.~\cite{Cross15}; however, in our case, such a quantum learning advantage is achieved with classical data.

To corroborate and extend our analysis, numerical simulations are performed: For a given $n$, we prepare a set of inputs $\{\mathbf{x}_1, \mathbf{x}_2, \ldots, \mathbf{x}_M \}$ that is sampled randomly. For the given $\omega(\mathbf{x}_i)$ ($i=1,2,\ldots,M$), we evaluate $P_{C,Q}(\omega)$ by counting $10^5$ queries and identify their average value, i.e., $\frac{1}{10^5} \sum_{i=1}^{10^5} P_{C,Q}(\omega(\mathbf{x}_i))$. This process is repeated $\simeq 10^3$ times for different input sets to analyze $\overline{P}_{C,Q}(n)$ statistically. The data are generated from $n=8$ to $n=35$, assuming that $\eta_k$ and $\chi_k$ ($\forall k$) are drawn from ${\cal N}(\overline{\eta}, \Delta_\eta)$ and ${\cal N}(\overline{\chi}, \Delta_\chi)$, respectively. Here, we consider $\overline{\eta}=10^{-3}$ with $\Delta_\eta = 0.05\overline{\eta}$ and $\overline{\chi}=10^{-2}$ with $\Delta_\chi = 0.1\overline{\chi}$. In each simulation, $M$ is fixed to $100$. The obtained data agree well with our theoretical predictions (see Figs.~2(a) and 2(b) in the main manuscript).

\end{document}